\documentclass[11pt]{article}

\usepackage{popovstyle}
\usepackage{booktabs}
\usepackage{longtable}

\usepackage{booktabs}
\usepackage{longtable}
\usepackage{array}
\usepackage{multirow}
\usepackage{wrapfig}
\usepackage{float}
\usepackage{colortbl}
\usepackage{pdflscape}
\usepackage{tabularx}
\usepackage{xltabular}
\usepackage{threeparttable}
\usepackage{threeparttablex}
\usepackage[normalem]{ulem}
\usepackage{makecell}
\usepackage{xcolor}

\title{Same Text, Different Filename: Are LLM Judges Biased?}
\author{Sergey V. Popov\footnote{Cardiff University,
\href{mailto:popovs@cardiff.ac.uk}{\nolinkurl{popovs@cardiff.ac.uk}}.
The usual disclaimer applies. Studies were preregistered at (1)
\url{https://aspredicted.org/sq9mp2.pdf}, then updated at
\url{https://aspredicted.org/qj59r2.pdf} and
\url{https://aspredicted.org/vp8ez6.pdf} and (2)
\url{https://aspredicted.org/82v8c4.pdf}. LLMs helped me with coding and
preparing the pre-registrations.}}
\date{\today}

\begin{document}

\maketitle

\begin{abstract}
Yes. In 2026, leading economic journals started using Refine.ink in the
technical quality assessment of submissions before publication. I study
whether renaming a file changes the perception of the document's quality
by LLM judges. I find that three frontier LLMs score the same document
differently if you change the document's filename. A person-related
label scores higher than no label; there is no consistent preference of
own- or rival-based label, nor a penalty for an Asian name, but there is
a context-specific bias in assessments whose direction and size vary
with both document and label. With LLM-written documents, some
self-premium can be detected, but a significant context-dependent label
bias remains. Whether to combat it is a question of the costs of
evaluation errors.
\end{abstract}

% Keywords

\medskip

% JEL classification

\medskip

\newpage

\section{Introduction}\label{introduction}

In 2026, Large Language Models (LLMs) are used by scientists in
developing both the content and the text of their studies. Since 2026 EU
law \citep[Article 50, ][]{eu2024aiact} requires LLMs to mark synthetic
output in machine-readable form. Verification of these marks should
allow editors to detect LLM use, and if the journal does not allow it,
prosecute the violators
\citep[indeed, arxiv.org has a one-strike policy against unchecked LLM use,][]{stan2026arxivban}.

On the other hand, direct use of LLMs in processing structured text that
made it omnipresent in coding environments allowed easy proofreading of
academic papers, and one of the refereeing startups, Refine.ink, was
officially adopted by editorial boards of leading Economics journals as
an intermediary step of evaluation of manuscripts before publication
\citep{golub2026refinepartners}. The combination of facilitated
detectability of LLM use, use of LLMs for judgment calls about textual
quality, and competition between LLM models creates a sensitive point:
would an LLM underlying an automated assessing tool be able to assess
the document ignoring whether it was treated by the tool or underlying
LLM?

This bias does not have to be malicious: treatment of manuscripts by
LLMs means maximizing a value function subject to constraints, and
different LLMs differ in values and constraints, so an LLM that used one
approach will deem something that maximized a different approach
suboptimal from their own perspective; whereas a document that was
already treated by the assessing LLM will look close to optimal just
because it was assessed by that LLM
\citep{deutsch-etal-2022-limitations}. There is literature about LLM
bias; this paper differs by asking whether the difference in the
\emph{filename} would introduce a difference in assessment.

To find out, I take 4 documents that I wrote myself, convert to textual
Markdown format to minimize metadata embedded in more sophisticated
formats, and manipulate the filename before asking LLMs to assess the
quality of the document's writing. This is not the same target as the
target of verification tools such as Refine.ink, but writing quality
assessment is LLMs' core competence, and therefore any deviations we
detect would be informative about the dangers for more specialized
tasks.

I find a small label bias, but more importantly, I find a significant
document-specific bias: it's comparable to the persistent bias. This
motivates me to run another study on LLM-generated texts to see whether
this document-specific bias is an artifact of a small number of
human-written texts in my corpus. With 90 LLM-generated texts, I find
the same label bias in the same assessors. This context-specific bias
has not been studied before, and its impact is at least as big as the
permanent bias. This has implications for fighting bias in non-LLM
assessment: if humans exhibit context-specific bias in assessment, this
bias might not only conceal the persistent bias by making data noisier,
but also would not be affected by persistent-bias lowering techniques.

The literature review is followed by the experimental description. Then
studies' results are presented, which are then followed by the
discussion and conclusion.

\section{Existing Literature}\label{existing-literature}

The bias of anonymous assessment of living referees in Economics is well
documented. \cite{blank1991effects} argued for double-blind review in
times before working papers were posted online. \cite{huber2022nobel}
ran an experiment on referees; apparently, Nobel prize winner signing a
paper immediately improves its novelty, relevance, and clarity of
writing, even if you don't change the text. Formatting affects
perception: \cite{james2026does} find that LaTeX-formatted papers are
perceived as of better quality. To test for biases, experiments with
label manipulations are numerous: \cite{bertrand2004emily} manipulated
names on CVs and found that HRs respond to some names more than to
others \citep[][ provides a survey]{bertrand2017field}. This paper uses
the same approach to manipulate the filename without changing the text
to see if LLM's assessment of the quality of writing is different.
Replacing biased humans with unbiased computers is tantalising
\citep{pataranutaporn2025peer,king2026peerless,mizumoto2023exploring,russolatona2024lottery,thakkar2026randomized,kim2026icml}.
LLMs however are not unbiased.

The sensitivity of LLMs to manipulations of texts and prompts is well
researched
\citep{lior2026comparingframingeffecthumans, robinson2025framinggamecontextshapes}.
Bias towards human authors is detected by above studies and others
\citep{everyone2025prefer,sun2026label}. In that literature,
\cite{saraf2025label} run an experiment similar to mine except their
texts are LLM-generated, and the label is part of their call to the LLM
judge; \cite{germani2025source} do the same thing, but also find bias
against Chinese-labelled sources, and one of my hypotheses tests that
bias \citep[see also][]{nghiem2024doctorlin}.
\cite{abolghasemi2025attribution} uses texts produced by humans and
LLMs, and submits authorship labels as part of the text tag. In my first
experiment, I was the author of all texts, so the difference between
different texts is not from which LLM it came, but only from any bias
that LLMs develop from seeing the document's filename; I detect some
self-preference in my second study. \cite{rospocher2026authormetadata}
shows that author identity (bibliometrics, institution tier) affects LLM
judge decisions.

Papers above, as others
\citep{deutsch-etal-2022-limitations,wataoka2024self, chae2026labels}
do, find LLM bias towards their own writing, but I show that some bias,
albeit less pronounced and with ambiguous direction, remains even if the
true writing is not of their own: it's not just about how things are
written, it's about the label. \cite{panickssery2024llm} confirm that
LLMs could recognize their own writing even before the watermark was
introduced as a response to EU regulation. My label exploration study is
different from these because it submits human-written text (written by
me), ruling out the \cite{deutsch-etal-2022-limitations} explanation
that LLMs demonstrate self-bias by design.

\section{Experiment}\label{experiment}

\textbf{Pipeline} I use 4 English language texts that were not available
online prior to this experiment that I wrote at different times: one
overview of the junior Welsh chess scene (Chess at 1167 words), one
assessment for a teaching certificate (Assignment at 2915 words), one
sample answer to an exam question (Cournot at 340 words), and one
segment from an unwritten public-audience book (Book at 1156 words). I
saved their MD5 and SHA256 hashes after removing identifying
information. These documents are available for replication by request.
They are saved as \texttt{.md} files to minimise the influence of file
tags and other excessive information of more sophisticated formats.

I use 3 widely used LLMs to evaluate these texts for the quality of
writing. I took these texts in particular because they are not very
good; I need to avoid the ceiling effect. I ask each LLM

\begin{verbatim}
Evaluate the attached document for quality of writing on a scale from 0 to 10. Use one
decimal place. Do not elaborate, just return the number.
\end{verbatim}

Each LLM separately assessed 6 versions of the file with identical
content but different filenames:

\begin{verbatim}
  Document.md                     Document_from_Sergey.md
  Document_from_Chen.md           Document_from_ChatGPT.md
  Document_from_Claude.md         Document_from_Gemini.md
\end{verbatim}

I use OpenRouter to access LLMs, which proved cheaper than buying
subscriptions. Anthropic Claude designed my R code for API calls.

\textbf{Hypotheses} As registered at
\href{https://aspredicted.org/vp8ez6.pdf}{AsPredicted}, I test 5
hypotheses:

\begin{itemize}
    \item[H1]  Documents labelled as written by an LLM (ChatGPT, Claude, or Gemini) receive lower scores, on average, than the same documents with a neutral or Sergey anchor.
    \item[H2] For each evaluating model, a document labelled as written by that model's own family receives a higher score than the same document labelled as written by either rival model.
    \item[H3] For the same evaluating models, documents labelled as written by a rival model receive lower scores than the same documents with a neutral filename.
    \item[H4 and H4b] Documents labelled with an Asian-coded human name receive lower scores on average than the same documents labelled with a non-Asian human or neutral filename. As a secondary comparison, I also report an Asian-coded label (Chen) against non-Asian (Sergey).
\end{itemize}

To test these hypotheses, I run an OLS regression of scores reported by
models on dummy variables of the used document and the used label
(neutral being the reference category), for each judge model separately.
Two-sided tests help me evaluate my hypotheses.

\textbf{Pilot} I used a free Ling 3.0 Flash VL
\texttt{inclusionai/ling-3.0-flash-vl:free} model via OpenRouter which
allowed 484 calls before my daily cap triggered (so the last call did
not return a result). Since the model is from neither of my 3 providers,
this allows me to estimate power needed for my human-vs-LLM evaluation.
The preliminary estimates showed H1's effect of \(-0.171\) with
\(0.050\) standard error and p-value of \(0.0006\); the model punishes
LLM label against human label by \(0.239\) (SE \(0.066\), p-value
\(0.0003\)). These effects are small compared to \cite{saraf2025label}:
they estimate shifts of \(\approx 12\) percentage points, if converted
to a comparable 0--10 scale. The residual standard deviation of
\(0.536\) suggests that if I want to detect a \(0.1\) size effect with a
power of \(90\%\), I need \(63\) repetitions to test H1, \(114\)
repetitions\footnote{The formula I used is
  \(N=\frac{k\sigma^2(z_{1-\alpha/2}+z_{1-\beta})^2}{\Delta^2}\), where
  \(\Delta=0.1\), \(\sigma=0.536\), \(\alpha=0.05\), power
  \(1-\beta=0.9\), and
  \(k=3\left(\frac{1}{3}\right)^2+2\left(\frac{1}{2}\right)^2\) for
  testing H1 vs \(k=1+2\left(\frac{1}{2}\right)^2\) to test H2-H4 (this
  depends on the weights of the coefficients in the test); divide by 4
  to obtain the number of repetitions.} for H2-H4, and 152 for H4b (its
\(k\) is 2). I therefore am running 152 requests per judge per document
per label.

\section{Human-written Texts}\label{human-written-texts}

After the ethics approval of Cardiff Business School's SREC panel was
obtained (\#3355), on October 27, 2026, I was able to run the code on
\texttt{anthropic/claude-sonnet-5.5}, \texttt{openai/gpt-6.1-sol}, and
\texttt{google/gemini-4-argon}, three models of similar capability from
different providers 152 repetitions per document-label yields 3648 calls
per model, or 10944 observations. Order of documents and labels is
randomized in a saved schedule. Every call is an independent
conversation, temperature is default and since no reasoning is recorded
or requested, max tokens is capped at 1000 (never reached). Routing was
managed by OpenRouter; Gemini 4 Argon got split between two Google
routes, with no statistically significant differences in outcomes. The
cost was about USD 36. The first run, executed in parallel, led to 5
problems in the data due to attempts to write in the same file
simultaneously, so I reran the failed 5 trials. The code and outcomes
can be obtained from the repository:
\url{https://github.com/sergeyvpopov/TextEval}.

Unfortunately, OpenRouter cuts off the filenames of uploaded files,
which made me submit the filename as part of the text. Another approach
would be to upload a .pdf and send a link for processing by the LLM, but
.pdfs have more metadata, so I decided to use the filename in text
instead. All 10944 replies were valid, realised residual SDs were far
below the pilot's, so I include the minimum detectable effect at 80\%
power (which I aimed to be \(0.1\) in power calculations).

All 10,944 marks lie between \(2.5\) and \(6.8\), so the models use less
than half of the 0--10 scale, and each uses a narrower band: GPT-6.1 Sol
uses {[}4.6,6.5{]}, Claude Sonnet 5.5 uses {[}2.5,6.5{]} and Gemini 4
Argon uses {[}4.2,6.8{]}. Against this, a typical label effect of
0.1--0.2 points is 1--2\% of the scale, but it is not small relative to
the spread of the marks. Within one document and label, 95\% of single
calls lie within \([-0.32,0.28]\) of the cell mean for Sol,
\([-0.52, +0.72]\) for Sonnet and \([-0.53, +0.50]\) for Argon. The
largest label differences within a single document reach \(0.39\) points
for Sol (Book), \(0.85\) for Sonnet (Book) and \(0.68\) for Argon
(Book), which for Sonnet is about 44\% of the gap between its highest
and lowest document means (\(1.9\) points). Marks are also coarse:
Sonnet's three most frequent values (\(3.5\), \(4.5\) and \(5.8\))
account for 76\% of its replies, and an average bias represents a change
in probabilities, not a shift in values.

\begin{table}[ht]
\centering
\caption{Registered contrasts by model; fixed effects for documents. MDE$_{80}$ is the minimum detectable effect at 80\% power.}
\label{tab:hypotheses}

\begin{tabular}{llrrlrr}
\toprule
Hypothesis & Model & Estimate & SE & 95\% CI & MDE$_{80}$ & $p$ (Holm)\\
\midrule
H1 & openai & -0.092 & 0.006 & {}[-0.104, -0.081] & 0.019 & $<$0.001\\
H1 & anthropic & 0.123 & 0.012 & {}[0.099, 0.147] & 0.039 & $<$0.001\\
H1 & google & -0.203 & 0.010 & {}[-0.224, -0.183] & 0.034 & $<$0.001\\
H2 & openai & -0.037 & 0.009 & {}[-0.054, -0.020] & 0.028 & $<$0.001\\
H2 & anthropic & -0.099 & 0.020 & {}[-0.137, -0.060] & 0.064 & $<$0.001\\
H2 & google & 0.046 & 0.014 & {}[0.019, 0.073] & 0.045 & $<$0.001\\
H3 & openai & -0.021 & 0.008 & {}[-0.037, -0.006] & 0.026 & 0.007\\
H3 & anthropic & 0.279 & 0.015 & {}[0.249, 0.309] & 0.050 & $<$0.001\\
H3 & google & -0.175 & 0.014 & {}[-0.203, -0.147] & 0.046 & $<$0.001\\
H4 & openai & 0.058 & 0.006 & {}[0.045, 0.070] & 0.021 & $<$0.001\\
H4 & anthropic & 0.073 & 0.015 & {}[0.043, 0.103] & 0.049 & $<$0.001\\
H4 & google & -0.010 & 0.014 & {}[-0.037, 0.018] & 0.046 & 0.492\\
H4b & openai & -0.001 & 0.007 & {}[-0.015, 0.014] & 0.024 & 0.911\\
H4b & anthropic & -0.050 & 0.018 & {}[-0.085, -0.016] & 0.057 & 0.008\\
H4b & google & -0.053 & 0.016 & {}[-0.085, -0.022] & 0.052 & 0.003\\
\bottomrule
\end{tabular} 
\end{table}

Estimation is a model-specific OLS regression as preregistered, with
document fixed effects and HC3 standard errors (can't cluster because
very few documents). I used Holm correction for multiple testing, with
\(\alpha=0.05\). The hypotheses are linear tests of dummy variable
coefficients.

\textbf{H1 (LLM-label penalty)} LLM penalty is visible but not uniform;
Anthropic on average prefers LLM-labelled output to human-labelled
output. The findings are consistent with expected; the impact is not
large, 0.1--0.2 on a 10-point scale on average. This is smaller than
what \cite{saraf2025label} finds.

\textbf{H2 (own-family premium)} There is no consistent own-family
effect. Argon shows self-preference (+0.046), others show small
self-deprecation. It could be driven by the fact that my documents are
not really written by LLMs; I did not test the bias on LLM-generated
documents.

\textbf{H3 (rival penalty)} The OpenAI model punishes other models; the
Anthropic model praises them, beyond what could be found due to
self-deprecation; the Google model punishes them; the effects are not
large. This could be driven by the fact that the documents are
human-generated.

\textbf{H4 (Chen vs Sergey+Neutral) and H4b (Chen vs Sergey)} Anti-Asian
bias of \cite{germani2025source} is borderline visible; compared to
Sergey label, Chen obtains about 0.05 lower score. There are probably
other biases (gender, specific names, etc) at play, and future research
can explore these venues. As an exploratory analysis, I also run a sign
test: 12 out of 13 significant tests remain significant with the same
sign; 0 out of 2 insignificant tests become significant.

\begin{table}
\centering
\caption{Exploratory analysis by document; differences in documents affect the biases studied by the hypotheses. Star denotes a 1\% significant effect, unadjusted for multiple testing; brackets contain HC3 standard errors.}
\label{tab:hyp_by_docs}

\begin{tabular}{llrrrrr}
\toprule
Model & Document & H1 & H2 & H3 & H4 & H4b\\
\midrule
GPT-6.1 Sol & Assignment & -0.07* & 0.03 & -0.03* & 0.11* & 0.07*\\
 &  & (0.01) & (0.01) & (0.01) & (0.01) & (0.01)\\
 & Book & -0.23* & -0.14* & -0.12* & -0.02 & -0.08*\\
 &  & (0.01) & (0.02) & (0.02) & (0.02) & (0.02)\\
 & Chess & -0.05* & -0.02 & 0.01 & 0.04* & -0.01\\
 &  & (0.01) & (0.01) & (0.01) & (0.01) & (0.01)\\
 & Cournot & -0.02 & -0.01 & 0.05* & 0.10* & 0.03\\
 &  & (0.01) & (0.02) & (0.02) & (0.01) & (0.02)\\
\midrule
Claude Sonnet 5.5 & Assignment & -0.15* & -0.09* & -0.11* & -0.20* & -0.21*\\
 &  & (0.01) & (0.01) & (0.02) & (0.01) & (0.02)\\
 & Book & 0.23* & -0.56* & 0.52* & 0.17* & 0.07\\
 &  & (0.02) & (0.02) & (0.02) & (0.04) & (0.05)\\
 & Chess & 0.19* & -0.49* & 0.77* & 0.14* & -0.28*\\
 &  & (0.01) & (0.03) & (0.02) & (0.02) & (0.02)\\
 & Cournot & 0.22* & 0.75* & -0.07* & 0.19* & 0.22*\\
 &  & (0.02) & (0.03) & (0.02) & (0.03) & (0.03)\\
\midrule
Gemini 4 Argon & Assignment & -0.20* & 0.07* & -0.17* & -0.06* & -0.11*\\
 &  & (0.02) & (0.03) & (0.03) & (0.02) & (0.02)\\
 & Book & -0.35* & 0.04 & -0.24* & -0.23* & -0.35*\\
 &  & (0.02) & (0.03) & (0.03) & (0.03) & (0.03)\\
 & Chess & -0.07* & 0.03 & -0.01 & 0.08* & 0.01\\
 &  & (0.01) & (0.01) & (0.02) & (0.02) & (0.02)\\
 & Cournot & -0.20* & 0.05 & -0.28* & 0.17* & 0.24*\\
 &  & (0.03) & (0.04) & (0.04) & (0.03) & (0.04)\\
\bottomrule
\end{tabular} 
\end{table}

Documents are of different lengths (Assignment is longest, Cournot is
shortest), written at different times, and aimed at different audiences.
This makes marks for different texts different; but, as Table
\ref{tab:hyp_by_docs} shows, biases are not document-neutral. Own-family
effects seem to be unstable: depending on the document, Sonnet can
exhibit both self-deprecation (-0.56 for Book and -0.49 for Chess) and
self-aggrandizing (Cournot +0.75); Gemini's effect is borderline zero.
Note that for the H4 anti-Asian bias the zero overall bias of Table
\ref{tab:hypotheses} averages large biases for individual documents.

Like any study, this study has limitations. I have only 4 documents;
their advantage is that I know they were done without any LLM
assistance, and LLMs were not trained on them before the experiment. I
test 6 labels; they were chosen ad hoc, and a different label could
exhibit a significant difference. The filename had to be sent via
OpenRouter as a part of the prompt; the effect of uploading a file might
differ. Exploratory analysis (one model, one document) indicates that
uploading matters. Indeed, the Claude run with Cournot via OpenRouter
yielded a mark of 3.56 for no label with an SD of 0.22, and a mark of
4.25 for a Claude label with an SD of 0.42. A similar run via Claude
(Opus opening Sonnet instances) yielded different marks when the text
was provided as with OpenRouter due to differences in harness, tools and
context: 5.57 SD 0.23 (neutral label) and 5.58 SD 0.28 (Claude label).
The same run with uploaded files changed marks to 4.77 SD 0.43 (neutral)
and 4.49 SD 0.11 (Claude). The difference in label bias can be
attributed to reading via a tool instead of submitting as part of the
prompt, so biases are not neutral to the delivery method.

The LLMs I studied score differently identical documents when only the
implied filename changes. A file with a human-name label scores higher
than an unlabelled document. LLM-labelled models score lower than no
label for Sol, higher for Sonnet, and lower for Argon. I find no
consistent preference for oneself for any of three models; Asian-coded
name label does not seem to carry a large penalty compared to Sergey.

Nevertheless, biases remain. The biases seem most pronounced when
interacted with documents. For instance, Gemini finds little difference
between Chen and Sergey labels on average, but that decomposes into a
+0.24 bonus for Chen for the Cournot document averaged with -0.35
penalty when the Book document is analyzed. Biases are not
document-neutral, context and delivery of manuscripts matter too;
insignificance of the bias on average does not mean the bias is not
there on the document level. Whether to tolerate the bias is an economic
question, which requires knowledge of the costs of decision errors of
both kinds, for both decisionmakers and applicants.

\section{Bias Decomposition}\label{bias-decomposition}

To see how much bias is persistent, and how much is context-sensitive, I
pre-registered a second study on
\href{https://aspredicted.org/82v8c4.pdf}{AsPredicted \#321045} on
October 29, 2026, to generate 10 briefs in 3 genres and feed them to 3
LLMs to generate texts for analysis. This obtains 90 texts, which I then
label as above, plus two labels, \texttt{\_v2} and
\texttt{\_from\_Catherine}. The first study informed that LLMs are
relatively persistent in their estimates, so there is little statistical
value in running more than 12 evaluations per text. I ask:

\begin{itemize}
\item[Q1] Does a label effect exist? My previous study found it for human writing.
\item[Q2] How large is it? Study 1 found a persistent bias as large as Anthropic preference for rival labels of $+0.279$, and a persistent penalty in Google evaluation of rival labels of $-0.175$. Document-specific bias was even larger: Claude gave a $0.77$ higher evaluation to texts labeled by rivals when working with the Chess text, and Gemini penalised the Cournot text written by rivals by $-0.28$. This study adds another human name and a version label.
\item[Q3] What is the proportion of persistent bias across documents? The exploratory analysis of the previous study calculates the label bias conditional on the document, and then calculates the average bias across labels for each LLM judge. This yields a persistent bias SD across documents of 0.07 for Sol, 0.11 for Sonnet, and 0.11 for Gemini; the document-specific part is 0.06 for Sol, 0.22 for Sonnet, and 0.09 for Argon. This means persistent bias is 57\% of realised bias for Sol, 19\% for Sonnet, and 62\% for Argon. These were not very reliable based on 4 texts; this study generates more texts, but uses LLM generation for this, and asks how much these shares change. \end{itemize}

The preregistered hypotheses (all tests two-sided) are

\begin{itemize}
    \item[H5] self-preference: own family scored higher under neutral label, predicted positive;
    \item[H6] own label on own text bias: predicted positive;
    \item[H7] LLM penalty: LLM labels score lower than human name labels, predicted negative.
\end{itemize}

\begin{table}[ht]
\centering
\caption{Named persistent biases, Study 2 (H5--H7), by judge. Unit of analysis is the brief (30 briefs, 29 degrees of freedom); $p$ values are Holm-adjusted across the three judges within each hypothesis. H5: score a judge gives its own family’s text minus the score predicted from judge and generator effects fitted on the other six judge–generator pairs, under the neutral label; H6: own-family label minus the mean of the two rival LLM labels, on own-family text; H7: LLM labels minus human names. Briefs shows positive/negative contrasts. MDE$_{80}$ is the minimum detectable effect at 80\% power.}
\label{tab:named_biases}

\begin{tabular}{llrrlrrr}
\toprule
Hypothesis & Judge & Estimate & SE & 95\% CI & Briefs & MDE$_{80}$ & $p$ (Holm)\\
\midrule
H5 & GPT-6.1 Sol & 0.338 & 0.049 & {}[0.238, 0.437] & 29/1 & 0.165 & $<$0.001\\
 & Claude Sonnet 5.5 & -0.281 & 0.068 & {}[-0.419, -0.142] & 5/25 & 0.230 & $<$0.001\\
 & Gemini 4 Argon & 0.551 & 0.083 & {}[0.382, 0.721] & 30/0 & 0.281 & $<$0.001\\
\midrule
H6 & GPT-6.1 Sol & 0.018 & 0.006 & {}[0.005, 0.032] & 20/10 & 0.022 & 0.016\\
 & Claude Sonnet 5.5 & 0.019 & 0.024 & {}[-0.030, 0.067] & 13/16 & 0.081 & 0.436\\
 & Gemini 4 Argon & 0.157 & 0.022 & {}[0.112, 0.203] & 28/2 & 0.076 & $<$0.001\\
\midrule
H7 & GPT-6.1 Sol & -0.111 & 0.007 & {}[-0.125, -0.098] & 0/30 & 0.023 & $<$0.001\\
 & Claude Sonnet 5.5 & -0.120 & 0.018 & {}[-0.157, -0.082] & 4/26 & 0.062 & $<$0.001\\
 & Gemini 4 Argon & -0.127 & 0.010 & {}[-0.147, -0.108] & 0/30 & 0.033 & $<$0.001\\
\bottomrule
\end{tabular} 
\end{table}

The run cost USD 47.66, returned 25,920 valid responses, with 12
repetitions for every judge, and took 20 hours, much of which consisted
of delays between calls to not violate ToS of OpenRouter. The remaining
technical details were similar to the first study.

Self-preference is confirmed (acknowledging Sonnet's self-deprecation);
human preference is confirmed. To see the variation across documents
analogously to Table 2, observe Figure 1. Three columns represent
valuations by three judges. Each row shows evaluations with each label
(0 for no label, 1 for \texttt{\_v2}, 2--4 for OpenAI, Anthropic, and
Google, and 5--7 for Sergey, Chen, and Catherine). Documents are coded
by author (O for OpenAI, A for Anthropic and G for Google) followed by a
Cyrillic brief letter (see Appendix A for details of brief topics).
Label variation in a row shows the label effect; label variation across
documents shows variation across briefs and authors. The difference in
briefs is seen as difference in writing quality. The brief topics are
benign, with zero references to say crime to minimize the risk of
getting my account banned; a sensitive topic might be interpreted by an
LLM judge as a low (or high!) quality document. I don't study this, but
it is a clear venue for future researchers, and has clear implications
for LLM-based evaluations of the quality of mathematical analysis in
applied theory papers.

\begin{figure}[p]
\includegraphics[width=1\linewidth]{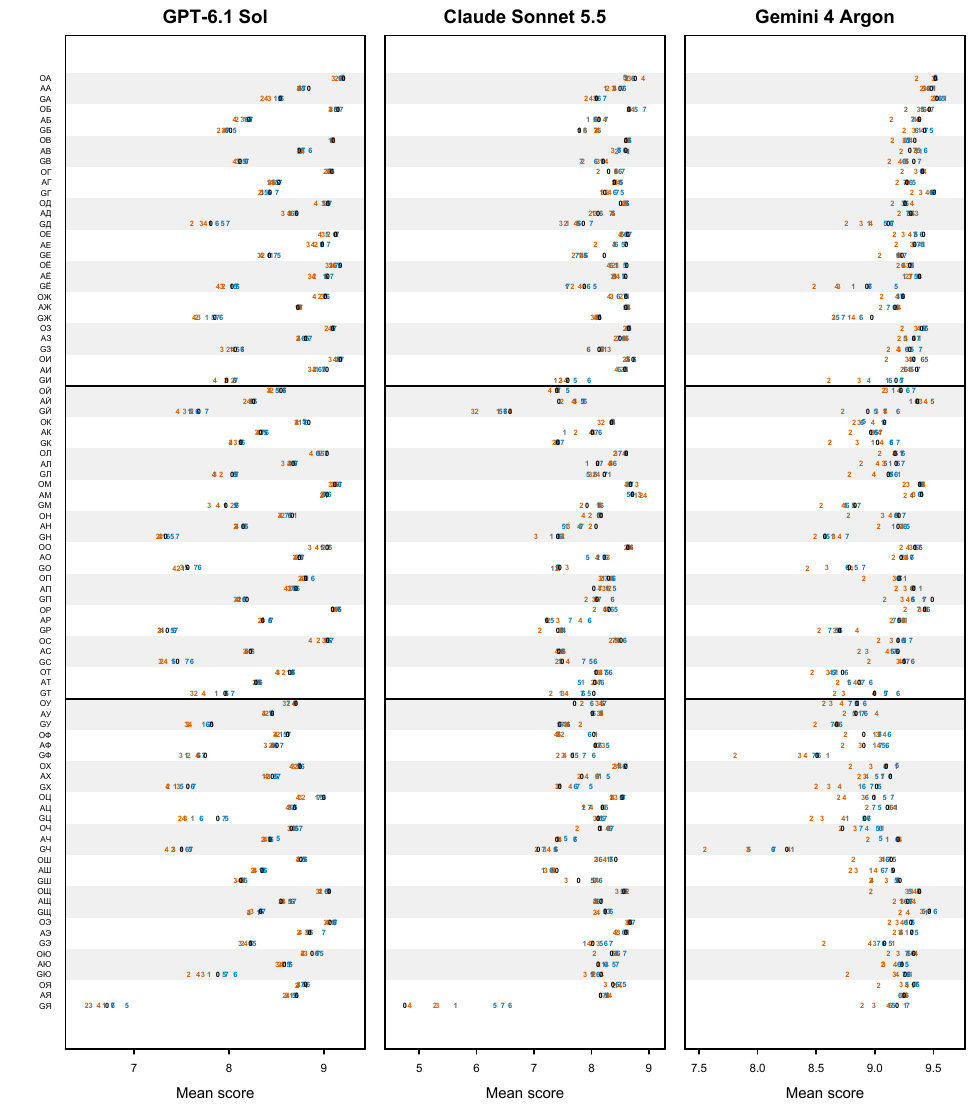} \caption{Scores for each of the 90 texts under each label, by judge (Study 2). Rows are texts: the first third are academic texts, the second general-audience essays, the last reports; lines separate. Each is coded by writer (O = OpenAI, A = Anthropic, G = Gemini) followed by a Cyrillic brief letter (key in the Table 5). Each digit is one label, placed at the text's mean score (over the 12 repetitions) under that label: 0 = neutral label (black); 1 = \_v2 (grey); 2, 3, 4 = ChatGPT, Claude, Gemini labels (orange); 5, 6, 7 = Sergey, Chen, Catherine (blue). Each panel spans that judge's lowest to highest text-by-label mean.}\label{fig:fig-doc-bias}
\end{figure}

To conclude, in Table 4 I report the share of the persistent bias
variance in total bias variance. LLM-written texts exhibit similar
patterns to human-written texts; it is not something specific to human
writing that drives context-specific biases, it's the interaction with
labels.

\begin{table}[H]
\centering
\caption{\label{tab:tab:perc_bias}Persistent share of label-related variance (percent). Corrected shares adjust for finite-repetition noise; 95\% cluster-bootstrap percentile intervals in brackets (2,000 resamples). Q3 columns use between-unit products (documents for Study 1, briefs for Study 2) as preregistered.}
\centering
\begin{tabular}[t]{lcccc}
\toprule
Judge & \shortstack{Human texts\\(corrected)} & \shortstack{LLM-written\\(corrected)} & \shortstack{Human texts\\(Q3)} & \shortstack{LLM-written\\(Q3)}\\
\midrule
GPT-6.1 Sol & 57.3\% [50.8, 94.9] & 63.6\% [58.9, 70.4] & 43.0\% [34.4, 93.2] & 63.1\% [58.4, 69.7]\\
Claude Sonnet 5.5 & 19.3\% [9.5, 87.7] & 17.6\% [12.9, 26.2] & -7.5\% [-20.7, 83.6] & 16.6\% [12.1, 25.3]\\
Gemini 4 Argon & 62.4\% [53.1, 92.1] & 62.7\% [57.8, 69.1] & 49.8\% [37.5, 89.5] & 62.3\% [57.1, 69.0]\\
\bottomrule
\end{tabular}
\end{table}

\section{Conclusion}\label{conclusion}

I find that filename labels introduce bias. The persistent bias is small
but detectable: LLMs prefer people over LLMs as authors, and absence of
informative label is often worse than both. However, besides the
persistent bias, there is context-specific bias, whose impact is
frequently stronger than the persistent part, and whose variability
masks the significance of the bias. I detect it in LLM judges; it could
be in human judges too.

The biases I document might decrease with time as models become more
sophisticated. To combat them, one could reduce the label effect to
no-label bias by scrambling any metadata before the analysis. The
invisible watermark, LLM verbiage (``load-bearing'', ``spine'',
em-dashes), and cadence are harder to get rid of, and they could be
interacting with documents too. Some of the bias can be averaged out by
using multiple LLMs multiple times and averaging the outcomes; such an
approach is unlikely to work with human judges. Future research can help
quantify the machinery behind the context-specific bias, which hopefully
will inform ways of combating it.

\begin{longtable}[t]{lll}
\caption{\label{tab:table-brief-key}Key to the brief codes in Figure 1. The confusing enumeration of briefs was my mistake when generating manuscripts; I decided against manual renaming.}\\
\toprule
Code & Brief & Description\\
\midrule
\includegraphics[height=1.5ex]{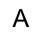} & B01CournotVsStackelberg & academic text: Cournot vs Stackelberg\\
\includegraphics[height=1.5ex]{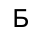} & B02Eignevalues & academic text on eigenvalues\\
\includegraphics[height=1.5ex]{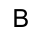} & B02InvisibleHand & academic text on Invisible Hand\\
\includegraphics[height=1.5ex]{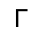} & B04Money & academic text on money neutrality\\
\includegraphics[height=1.5ex]{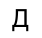} & B05BehaviouralFinance & academic text on behavioural finance\\
\addlinespace
\includegraphics[height=1.5ex]{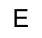} & B06WritingAPaper & academic text on paper writing\\
\includegraphics[height=1.5ex]{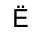} & B07WritingaSOP & academic text: PhD purpose statement\\
\includegraphics[height=1.5ex]{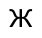} & B12Biases & academic text on LLM bias\\
\includegraphics[height=1.5ex]{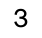} & B18Syllabus & academic text: machine-learning course outline\\
\includegraphics[height=1.5ex]{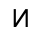} & B22InflationUnemployment & academic text: inflation and unemployment\\
\addlinespace
\includegraphics[height=1.5ex]{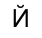} & B08Appeal & general essay: parking appeal letter\\
\includegraphics[height=1.5ex]{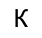} & B09Chess & general essay on UK chess\\
\includegraphics[height=1.5ex]{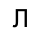} & B10Mars & general essay on Mars colony\\
\includegraphics[height=1.5ex]{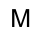} & B11LLMUse & general essay on LLM self-education\\
\includegraphics[height=1.5ex]{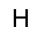} & B13ChessAd & general essay: chess for children\\
\addlinespace
\includegraphics[height=1.5ex]{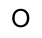} & B15SourcesOfCreativity & general essay on creativity sources\\
\includegraphics[height=1.5ex]{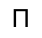} & B21Brexit & general essay on Brexit\\
\includegraphics[height=1.5ex]{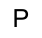} & B25Lovecraft & general essay on Lovecraft's pantheon\\
\includegraphics[height=1.5ex]{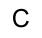} & B26Coffee & general essay: espresso versus jezve\\
\includegraphics[height=1.5ex]{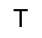} & B30Fields & general essay: employable European fields\\
\addlinespace
\includegraphics[height=1.5ex]{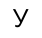} & B16RealEstate & report on Welsh real estate\\
\includegraphics[height=1.5ex]{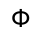} & B17Travel & report: Spanish scenic road trip\\
\includegraphics[height=1.5ex]{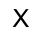} & B17Universities & report on German higher education\\
\includegraphics[height=1.5ex]{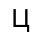} & B19Sleep & report on sleep deprivation\\
\includegraphics[height=1.5ex]{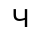} & B20CookingPlan & report: week of student meals\\
\addlinespace
\includegraphics[height=1.5ex]{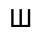} & B23Carcassonne & report: Carcassonne and similar games\\
\includegraphics[height=1.5ex]{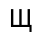} & B24Bardic & report: Welsh bards versus troubadours\\
\includegraphics[height=1.5ex]{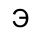} & B27Painting & report on Galileo's scientific role\\
\includegraphics[height=1.5ex]{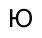} & B28Art & report on Basquiat's artistic role\\
\includegraphics[height=1.5ex]{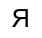} & B29Tales & report: Welsh tale tropes\\
\bottomrule
\end{longtable}

\bibliographystyle{plainnat}
\bibliography{references}

\end{document}